\documentclass{article}
\usepackage{graphicx} 
\usepackage{amsmath}
\usepackage{tabularx}
\usepackage{caption}
\usepackage{subcaption}
\usepackage{cite} 

\usepackage[margin=1in]{geometry}
\title{Painlev\'e Integrability,  Auto-B\"acklund Transformation and the exact solutions of (1+2) Kudryashov-Sinelshchikov (KS) equation.}

\author{
	Apeksha Patil$^{1}$,
	Amlan Kanti Halder$^{2}$,
	Rajeswari Seshadri$^{3}$\\[1ex]
	$^{1,3}$Department of Mathematics, Pondicherry University, Puducherry, India\\
	$^{2}$School of Sciences, Woxsen University, Hyderabad, India\\
	$^{1}$\texttt{patilapeksha2499@gmail.com}\\
	$^{2}$\texttt{amlanhalder1@gmail.com}\\
Corresponding Author: $^{3}$\texttt{seshadrirajeswari@gamil.com}
}
\date{}
\begin{document}
	\maketitle
	\section*{Abstract}
	In this research work, we consider a nonlinear fourth-order (1+2)-dimensional Kudryashov-Sinelshchikov (KS) equation which represents the wave propagation of pressures in liquids that contain gas bubbles. A direct Integrability of the KS equation is analysed using Painlev\'e Analysis with Singular Manifold Method (SMM).  With the help of WTC algorithm,  we show that the (1+2) KS equation is Painlev\'e integrable. Then by truncating the Painlev\'e expansion we obtain the Auto-B\"acklund Transformation (ABT). By taking suitable forms of Manifold, various exact solutions based on the obtained Auto-B\"acklund transformation are derived.  The consistancy check for these solutions are also performed. Representative solutions are presented in the from of 2D and 3D plots to understand the geometric perspective of the solutions.
	\newline
	\textbf{Keywords:} Kudryashov-Sinelshchikov (KS) equation, Painlev\'e Anaylsis of PDE, Singular Manifold Method, Auto-B\"acklund Transformation, Exact solutions.
	\section{Introduction}
	
	Nonlinear evolution equations arise naturally in many areas of applied mathematics and physics, where they describe wave propagation and other complex dynamical phenomena. They are widely used to model nonlinear wave motion in fluids, plasmas, optical fibers, and other continuous media. Unlike linear models, nonlinear equations capture essential physical effects such as wave steepening, wave interaction, and the formation of localized structures. One of the most remarkable outcomes of the interplay between nonlinearity and dispersion is the appearance of solitary waves.
	\smallskip
		
	Another key problem in nonlinear evolution equation study is the question of integrability. An equation is termed to be integrable when it contains certain special mathematical characteristics that enable it in the construction of exact solutions using analytical values. Integrable equations usually have soliton solutions, conservation laws and transformation structures that can be used to learn about the dynamics of the flow phenomena.
	\smallskip
	
	The Painlev\'e analysis is one of the existing approaches to investigating integrability. In the case of ordinary differential equations (ODEs), the main concept states that the general solution of an equation that is integrable, must not have movable critical singularities beyond poles. When this condition holds, then the ODE is said to possess the Painlev\'e property. This property is an adequate integrability property, but is not a necessary one.
	However, in the case of partial differential equations (PDEs), the idea is generalized to the action of solutions around moving singularity manifolds. Under such condition, the equation is said to have the Painlev\'e property which is the indication of a strong integrability. The research of Weiss, Tabor, and Carnevale introduced a systematic process of applying the Painlev\'e test to nonlinear partial differential equations, which became referred to as the WTC algorithm \cite{Weiss1983}. Their methodology was further developed and extended in later works, such as truncation methods \cite{Pickering1993}, the singular manifold method \cite{Estevez2005}, and comprehensive theoretical treatments of the subject \cite{Musette1999}.
	\smallskip
	
	The Auto-B\"acklund Transformation (ABT) is closely related to the Painlev\'e analysis. Transformation relations between solutions can be established by truncating the Painlev\'e series at some level. The resulting ABT gives a direct and effective method of producing new exact solutions from known ones.
	\smallskip
	
	The Painlev\'e analysis has been applied successfully to other higher-dimensional nonlinear evolution equations in recent years. As an example, the multi-soliton structures and the integrability properties of variable-coefficient equations were studied by Hao \cite{Hao2024}. However, Painlev\'e analysis and Auto-B\"acklund transformations were utilized to obtain analytical solutions of generalized Burgers equations and generalized Sakovich equations in the works of Zhou and co-authors and Singh and Ray \cite{Zhou2022,Singh2022}. These papers point to the usefulness of the Painlev\'e framework to display integrable characteristics and create exact solutions of nonlinear models.
	\smallskip

	 Kudryashov and Sinelshchikov \cite{Kudryashov2010,Kudryashov2012} proposed a nonlinear partial differential equation to model the propagation of pressure waves in liquids containing gas bubbles in 2010. After its introduction, the Kudryashov--Sinelshchikov (KS) equation has been widely recognized as an important model in nonlinear acoustics and the study of bubbly flows. The general $(1+3)$-dimensional form of the KS equation \cite{Ali2021} is written as
	\begin{equation}
	 d u_{yy} + e u_{zz}+	(u_t + \alpha u u_x + \gamma u_{xxx})_x = 0,
	\end{equation}
	where $u(t,x,y,z)$ denotes a physical variable such as density, pressure, or fluid velocity. The parameters  $d$, $\alpha$, $\gamma$, and $e$ represent the nonlinear and dispersive characteristics of the medium.
	\smallskip
	
	Since its introduction, the $(1+3)$-dimensional KS equation has been investigated using different analytical and symmetry-based approaches. Lie symmetry analysis, similarity reductions, and invariant solutions were studied in \cite{Yang2015,Ali2021,ElShiekh2022}, while bifurcation behavior and the influence of viscosity and heat transfer were analyzed in \cite{Feng2014,Seadawy2019}. In addition, several authors have derived exact solutions using diverse analytical techniques. Solitary wave solutions and conservation laws were reported in \cite{Chukkol2018,Inc2017,Rehman2021,Akram2021}, whereas complex wave phenomena such as lump, breather, rogue wave, and multi-soliton interactions were obtained in \cite{Tang2020,FengBilige2021,Kuo2021,Hu2021,Cr2022}. These studies demonstrate the rich mathematical structure and diverse nonlinear dynamics supported by the KS equation.
	\smallskip

	Later, Nuruzzaman \textit{et al.} \cite{nuruzzaman2023localized} reduced the $(1+3)$-dimensional KS equation to a $(1+2)$-dimensional form by imposing the transformation $z=x$, leading to
	\begin{equation}
	d u_{yy} + e u_{xx}+(u_t + \alpha u u_x + \gamma u_{xxx})_x  = 0.
	\end{equation}
	Although reduced in dimension, this model preserves the essential nonlinear and dispersive features of the original system and remains suitable for studying pressure wave propagation in liquid–gas mixtures. Using the Hirota bilinear method, Nuruzzaman \textit{et al.}\cite{nuruzzaman2023localized} constructed various families of solutions, including multi-soliton, breather, lump, and mixed interaction structures, and their graphical analysis revealed rich nonlinear wave interactions.
	\smallskip
	
	From the above discussion, it can be observed that for the $(1+2)$-dimensional KS equation, most of the available studies mainly focus on the Hirota bilinear method for constructing exact solutions. To the best of our knowledge, the integrability of this reduced model through Painlev\'e analysis of PDE has not yet been examined in detail.
	\smallskip
	
	In our previous work \cite{Preprint}, we investigated the $(1+2)$-dimensional KS equation using Lie symmetry analysis. We derived the symmetry generators, constructed similarity reductions, and obtained several invariant solutions. Further, with the help of multiplier approach we derived the conservation laws and hence, verified the conserved vectors. These results provided useful information about the structural properties of the equation. However, the question of integrability of this model still remains open.
	\subsection*{Objective of the present paper:}
	Motivated by this, in the present work we focus on studying the integrability of the $(1+2)$-dimensional KS equation by applying the direct Painlev\'e analysis of PDE by using WTC-algorithm. The advantage of applying this method is that if the PDE passes the Painlev\'e property it is said to be integrable in nature and the explicit solutions is obtained in Laurent series expansion around the singular manifolds. The paper is arranged in the following sections: 
	\smallskip
	
	In Section 2, we briefly describe the main steps of the Painlev\'e analysis of PDE based on the WTC algorithm and apply the procedure to the given partial differential equation. In Section 3, we truncate the Laurent expansion to construct the Auto-B\"acklund Transformation. Using an appropriate choice of singular manifold, we then derive new exact solutions of the equation. Finally, graphical representations of the obtained solutions are presented through two-dimensional and three-dimensional plots to illustrate their geometrical behavior.
	
	\section{Painlev\'e Analysis of PDE}
The Painlev\'e analysis of the partial differential equation is carried out using the singular manifold method (SMM), which follows the Weiss--Tabor--Carnevale (WTC) algorithm. If nonlinear partial differential equations(NLPDEs) passes painlev\'e test, the differential equation is said to be integrable in nature, whereas the Laurent series is truncated at a constant term to study about the closed-form solutions for the NLPDEs.
	\subsection{Steps in Painlev\'e analysis}
	Let us consider a nonlinear partial differential equations,
	\begin{equation}
		S(u,u_{t},u_{x},u_{y},u_{tt},u_{xx},\dots)=0.
	\end{equation}
	where $S$ is the polynomial in dependent variable $u=u(t,x,y)$ and $t,x,$ and $y$ are the independent variables.
	\newline
	The singular manifold \cite{Weiss1983,Estevez2005} for the above NLPDE is defined as 
	\begin{equation}
		\psi(t,x,y)=0.
	\end{equation}
	The Laurent series is consider as solution for the PDE(3) which is
	\begin{equation}
		u(t,x,y)=\psi^{\lambda}(t,x,y)\sum_{j=0}^{\infty}u_{j}(t,x,y)\psi^{j}(t,x,y),
	\end{equation}
	where $u_{j}$'s are the arbitrary functions of $(t,x,y)$ and $\lambda$ is negative parameter.
	\newline
	The Painlev\'e analysis of Eq.~(3) is performed as follows: 
	\begin{itemize}
		\item 	First, the leading-order exponent $\lambda$ is determined by substituting
		\begin{equation}
			u = \psi^{\lambda} u_{0},
		\end{equation}
		into Eq.~(3).
		\newline
			By equating the coefficient of the dominant term to zero, a nontrivial value known as leading-order coefficient $u_{0}$ is obtained. 
		This procedure may result in several possible dominant balances, corresponding to different solution branches.
		\item Next, to determine the resonance values, the expansion
		\begin{equation}
			u = u_{0} \psi^{\lambda} + \sum_{j=1}^{\infty} u_{j} \psi^{j+\lambda},
		\end{equation}
		is substituted into Eq.~(3). 
		The resulting indicial equation leads to a polynomial $Q(j)$ in the
		resonance parameter $j$. By equating the polynomial $Q(j)$ to zero,
		we obtain the resonance values. The obtained values of $j$ represent
		the resonance positions, which indicate the levels in the Laurent
		expansion at which arbitrary functions may enter the solution.
		\item The Laurent expansion given in Eq.~(5) is truncated at the highest resonance level $j = j_{\max}$, which results in the finite series
		\begin{equation}
			u = \psi^{\lambda} \sum_{j=0}^{j_{\max}} u_{j} \psi^{j}.
		\end{equation}
		The coefficients $u_{j}$ are then obtained step by step by inserting this truncated expansion into the governing equation and equating coefficients of like powers of $\psi$.
		
		\item For the Painlev\'e test to be satisfied, the coefficients $u_{j}$ associated with the resonance values must remain arbitrary functions. 
		\newline
		If this requirement holds at every resonance level, then Eq.~(3) is considered to satisfy the Painlev\'e test and possesses the Painlev\'e property. Hence, the given PDE is integrable in nature.
%
	\end{itemize}
 \subsection{Painlev\'e Integrability Analysis of the $(1+2)$-Dimensional KS Equation}
 Consider the (1+2)-Dimensional Kudryashov-Sinelshchikov (KS) equation given as:
 \begin{equation}
 	d u_{yy} + e u_{xx}+(u_t + \alpha u u_x + \gamma u_{xxx})_x  = 0.
 \end{equation}
 We assume that the solution form of the (1+2) KS equation around the singular manifold $\psi(t,x,y)=0$ as the Laurent series exapansion is defined as:
 \begin{equation}
 	u(t,x,y)=\psi^{\lambda}(t,x,y)\sum_{j=0}^{\infty}u_{j}(t,x,y)\psi^{j}(t,x,y),
 \end{equation}
 where $\lambda$ is a negative integer and $u(t,x,y), u_{j}(t,x,y)$ such that $u_{0}(t,x,y) \ne 0$ are the analytic functions in $t,x,$ and $y$.
 \newline
 \textbf{Step 1:} Substituting $u = \psi^{\lambda} u_{0}$ into the Eq.~(9), we determine the value of $\lambda$ and $u_{0}$ given as:
 \begin{equation}
 	\lambda=-2, \quad
 	u_0(t,x,y)= -\frac{12 \gamma  \psi _{\text{, }x}{}^2}{\alpha }.
 \end{equation}
 \textbf{Step 2:} Next step is to determine the resonance values, therefore we consider the expansion given by Eq.(5) and substituting the above obtained values in the expansion we obtained,
 
 	\begin{equation}
 		u = u_{0} \psi^{-2} + \sum_{j=1}^{\infty} u_{j} \psi^{j-2},
 	\end{equation}
Substituting above series expansion into the governing PDE(9), we gets the recursion relation. By setting the coeffecients of $\psi^{j-6}$ to zero , we get the polynomial $Q$ in $j$ of degree four, which is equated to zero
\begin{equation}
	Q(j)=\gamma j^{4} \psi_x^{4}
	-14 \gamma j^{3} \psi_x^{4}
	+59 \gamma j^{2} \psi_x^{4}
	-46 \gamma j \psi_x^{4}
	-120 \gamma \psi_x^{4}=0
\end{equation}
By solving above polynomial $Q(j)$ we get values of $-1,4,5,$ and $6$.
Therefore the resonance values are $-1,4,5,$ and $6$.
\newline
\textbf{Step 3:} From the resonance calculation, the largest resonance value is found to be $j_{max}=6$. Therefore, the Laurent series expansion given in Eq.~(12) is truncated up to the corresponding term as
\begin{equation}
u=\frac{u_0}{\psi ^2}+\frac{u_1}{\psi }+u_2+u_3 \psi+u_4 \psi ^2+u_5 \psi ^3+u_6 \psi ^4.
\end{equation}
Substituting this truncated expansion into the governing equation (9) and equating the coefficients of like powers of $\psi$ to zero, we obtain the following expressions:
\begin{equation}
	\begin{aligned}
		u_0 &= -\frac{12\gamma \psi_x^{2}}{\alpha}, \\[1ex]
		u_1 &= \frac{12\gamma \psi_{xx}}{\alpha}, \\[1ex]
		u_2 &= -\frac{d \psi_y^{2}}{\alpha \psi_x^{2}}
		-\frac{\psi_t}{\alpha \psi_x}
		+\frac{3\gamma \psi_{xx}^{2}}{\alpha \psi_x^{2}}
		-\frac{4\gamma \psi_{xxx}}{\alpha \psi_x}
		-\frac{e}{\alpha}, \\[1ex]
		u_3 &= -\frac{d \psi_{xx} \psi_y^{2}}{\alpha \psi_x^{4}}
		+\frac{d \psi_{yy}}{\alpha \psi_x^{2}}
		-\frac{\psi_t \psi_{xx}}{\alpha \psi_x^{3}}
		+\frac{\psi_{tx}}{\alpha \psi_x^{2}} 
		+\frac{3\gamma \psi_{xx}^{3}}{\alpha \psi_x^{4}}
		-\frac{4\gamma \psi_{xxx} \psi_{xx}}{\alpha \psi_x^{3}}
		+\frac{\gamma \psi_{xxxx}}{\alpha \psi_x^{2}} .
	\end{aligned}
\end{equation}
\textbf{Step 4:} The resonance value $j=-1$ corresponds to the arbitrariness of the singular manifold $\psi(t,x,y)=0$. 
Collecting the coefficients of $\psi^{-2}$, $\psi^{-1}$, and $\psi^{0}$ gives zero identically, which confirms the consistency of the expansion. Since the terms $u_4$, $u_5$, and $u_6$ do not appear in the compatibility conditions, they remain arbitrary functions. Similarily, $u_{7}, u_{8}, u_{9},\dots$ can be derived by using the above values. Hence, all compatibility conditions are satisfied, and Eq.~(9) possesses the Painlev\'e property. Hence, confirming its integrability in nature.
 \section{Construction of the Auto-B\"acklund Transformation and Exact Solutions}
 To derive the Auto-B\"acklund Transformation (ABT) \cite{Zhou2022,Singh2022}, we use the truncated Painlev\'e expansion approach. By making use of Painlev\'e property, the Laurent expansion presented in Eq.~(10) is truncated at the constant level in the neighborhood of the singular manifold.
 \newline
 We obtained leading order exponent $\lambda=-2$, therefore we truncated the Laurent series at $u_{2}$. This truncation procedure yields the required ABT, which can be written as follows:
%
%
 \begin{equation}
 	u=\frac{u_0}{\psi ^2}+\frac{u_1}{\psi }+u_2
 \end{equation}
 Substituting the above Eq.~(16) into the governing PDE(9) and collecting the coefficients of power of $\psi$ and equating it to zero we have the following equations:
 \begin{equation}
 	\psi^{-6}:\quad
 	10\alpha u_0^{2}\psi_x^{2}
 	+120\gamma u_0 \psi_x^{4}=0 .
 \end{equation}
 \begin{equation}
 	\begin{aligned}
 		\psi^{-5}:\quad
 		&12\alpha u_0 u_1 \psi_x^{2}
 		-8\alpha u_0 u_{0,x}\psi_x
 		-2\alpha u_0^{2}\psi_{xx} \\
 		&\quad
 		+24\gamma u_1 \psi_x^{4}
 		-96\gamma u_{0,x}\psi_x^{3}
 		-144\gamma u_0 \psi_{xx}\psi_x^{2}=0 .
 	\end{aligned}
 \end{equation}
 \begin{equation}
 	\begin{aligned}
 		\psi^{-4}:\quad
 		&6d u_0 \psi_y^{2}
 		+6e u_0 \psi_x^{2}
 		+6u_0 \psi_t \psi_x
 		+3\alpha u_1^{2}\psi_x^{2}
 		+6\alpha u_0 u_2 \psi_x^{2} \\
 		&-6\alpha u_1 u_{0,x}\psi_x
 		-6\alpha u_0 u_{1,x}\psi_x
 		-3\alpha u_0 u_1 \psi_{xx}
 		+\alpha u_{0,x}^{2}
 		+\alpha u_0 u_{0,xx} \\
 		&-24\gamma u_{1,x}\psi_x^{3}
 		-36\gamma u_1 \psi_{xx}\psi_x^{2}
 		+36\gamma u_{0,xx}\psi_x^{2}
 		+72\gamma u_{0,x}\psi_{xx}\psi_x \\
 		&+24\gamma u_0 \psi_{xxx}\psi_x
 		+18\gamma u_0 \psi_{xx}^{2}=0 ,
 	\end{aligned}
 \end{equation}
 \begin{equation}
 	\begin{aligned}
 		\psi^{-3}:\quad
 		&2d u_1 \psi_y^{2}
 		-4d u_{0,y}\psi_y
 		-2d u_0 \psi_{yy}
 		+2e u_1 \psi_x^{2}
 		-4e u_{0,x}\psi_x
 		-2e u_0 \psi_{xx} \\
 		&+2u_1 \psi_t \psi_x
 		-2\psi_t u_{0,x}
 		-2u_{0,t}\psi_x
 		-2u_0 \psi_{tx}
 		-\alpha u_1^{2}\psi_{xx}
 		+2\alpha u_2 u_1 \psi_x^{2} \\
 		&-4\alpha u_1 u_{1,x}\psi_x
 		-4\alpha u_2 u_{0,x}\psi_x
 		-4\alpha u_0 u_{2,x}\psi_x
 		-2\alpha u_0 u_2 \psi_{xx}
 		+\alpha u_1 u_{0,xx} \\
 		&+2\alpha u_{0,x}u_{1,x}
 		+\alpha u_0 u_{1,xx}
 		+6\gamma u_1 \psi_{xx}^{2}
 		+8\gamma u_1 \psi_x \psi_{xxx} \\
 		&+24\gamma u_{1,x}\psi_x\psi_{xx}
 		-12\gamma u_{0,xx}\psi_{xx}
 		+12\gamma u_{1,xx}\psi_x^{2}
 		-8\gamma u_{0,x}\psi_{xxx} \\
 		&-8\gamma u_{0,xxx}\psi_x
 		-2\gamma u_0 \psi_{xxxx}=0 .
 	\end{aligned}
 \end{equation}
 \begin{equation}
 	\begin{aligned}
 		\psi^{-2}:\quad
 		&-2d u_{1,y}\psi_y
 		-d u_1 \psi_{yy}
 		+d u_{0,yy}
 		-2e u_{1,x}\psi_x
 		-e u_1 \psi_{xx}
 		+e u_{0,xx} \\
 		&-\psi_t u_{1,x}
 		-u_{1,t}\psi_x
 		-u_1 \psi_{tx}
 		+u_{0,tx}
 		-2\alpha u_2 u_{1,x}\psi_x
 		-2\alpha u_1 u_{2,x}\psi_x \\
 		&-\alpha u_1 u_2 \psi_{xx}
 		+\alpha u_{1,x}^{2}
 		+2\alpha u_{0,x}u_{2,x}
 		+\alpha u_2 u_{0,xx}
 		+\alpha u_1 u_{1,xx} \\
 		&+\alpha u_0 u_{2,xx}
 		-4\gamma u_{1,x}\psi_{xxx}
 		-6\gamma u_{1,xx}\psi_{xx}
 		-4\gamma u_{1,xxx}\psi_x
 		-\gamma u_1 \psi_{xxxx}
 		+\gamma u_{0,xxxx}=0 .
 	\end{aligned}
 \end{equation}
 \begin{equation}
 	\psi^{-1}:\quad
 	d u_{1,yy}
 	+e u_{1,xx}
 	+u_{1,tx}
 	+2\alpha u_{1,x}u_{2,x}
 	+\alpha u_2 u_{1,xx}
 	+\alpha u_1 u_{2,xx}
 	+\gamma u_{1,xxxx}=0 .
 \end{equation}
 \begin{equation}
 	\psi^{0}:\quad
 	d u_{2,yy}
 	+e u_{2,xx}
 	+u_{2,tx}
 	+\alpha u_{2,x}^{2}
 	+\alpha u_2 u_{2,xx}
 	+\gamma u_{2,xxxx}=0 .
 \end{equation}
 By solving Eq.~(17) and Eq.~(18) we obtain the values of $u_{0}$ and $u_{1}$ given as,
 \begin{equation}
 	u_0=-\frac{12\gamma \psi_x^{2}}{\alpha} \qquad 
 		u_1=\frac{12\gamma \psi_{xx}}{\alpha}.
 \end{equation}
 with the help of Eq.~(24), Eq.~(19-23) becomes,
 \begin{equation}
 	\resizebox{\textwidth}{!}{$
 		\begin{aligned}
 			&-72\gamma \psi_x^{4} u_2
 			-\frac{72\gamma d \psi_x^{2}\psi_y^{2}}{\alpha}
 			-\frac{72\gamma e \psi_x^{4}}{\alpha}
 			-\frac{72\gamma \psi_t \psi_x^{3}}{\alpha}
 			-\frac{288\gamma^{2} \psi_{xxx}\psi_x^{3}}{\alpha}
 			+\frac{216\gamma^{2} \psi_{xx}^{2}\psi_x^{2}}{\alpha}=0,\\
 			&144\gamma \psi_{xx}\psi_x^{2} u_2
 			+\frac{24\gamma d \psi_x^{2}\psi_{yy}}{\alpha}
 			+\frac{96\gamma d \psi_x\psi_y\psi_{xy}}{\alpha}
 			+\frac{24\gamma d \psi_{xx}\psi_y^{2}}{\alpha}
 			+\frac{144\gamma e \psi_{xx}\psi_x^{2}}{\alpha}
 			+\frac{72\gamma \psi_x^{2}\psi_{tx}}{\alpha}
 			+\frac{72\gamma \psi_t \psi_{xx}\psi_x}{\alpha}\\
 			&\quad
 			+48\gamma u_{2,x}\psi_x^{3}
 			+\frac{216\gamma^{2} \psi_{xxxx}\psi_x^{2}}{\alpha}
 			-\frac{72\gamma^{2} \psi_{xx}^{3}}{\alpha}=0,\\
 			&-36\gamma \psi_{xx}^{2} u_2
 			-48\gamma \psi_x\psi_{xxx} u_2
 			-\frac{24\gamma d \psi_{xy}^{2}}{\alpha}
 			-\frac{24\gamma d \psi_x\psi_{xyy}}{\alpha}
 			-\frac{12\gamma d \psi_{xx}\psi_{yy}}{\alpha}
 			-\frac{24\gamma d \psi_y\psi_{xxy}}{\alpha}
 			-\frac{36\gamma e \psi_{xx}^{2}}{\alpha}
 			-\frac{48\gamma e \psi_x\psi_{xxx}}{\alpha}\\
 			&\quad
 			-\frac{12\gamma \psi_t\psi_{xxx}}{\alpha}
 			-\frac{36\gamma \psi_{xx}\psi_{tx}}{\alpha}
 			-\frac{36\gamma \psi_x\psi_{txx}}{\alpha}
 			-72\gamma u_{2,x}\psi_x\psi_{xx}
 			-12\gamma u_{2,xx}\psi_x^{2}
 			+\frac{24\gamma^{2} \psi_{xxx}^{2}}{\alpha}
 			-\frac{36\gamma^{2} \psi_{xx}\psi_{xxxx}}{\alpha}
 			-\frac{72\gamma^{2} \psi_x\psi_{xxxxx}}{\alpha}=0,\\
 			&12\gamma \psi_{xxxx} u_2
 			+\frac{12\gamma d \psi_{xxyy}}{\alpha}
 			+\frac{12\gamma e \psi_{xxxx}}{\alpha}
 			+\frac{12\gamma \psi_{txxx}}{\alpha}
 			+12\gamma u_{2,xx}\psi_{xx}
 			+24\gamma u_{2,x}\psi_{xxx}
 			+\frac{12\gamma^{2} \psi_{xxxxxx}}{\alpha}=0,\\
 			&\alpha u_{2,xx}u_2
 			+d u_{2,yy}
 			+e u_{2,xx}
 			+u_{2,tx}
 			+\alpha u_{2,x}^{2}
 			+\gamma u_{2,xxxx}=0
 		\end{aligned}
 		$}.
 \end{equation}
It is evident that Equations.~(19)--(23) are satisfied if and only if $\psi_{x} \neq 0$.
Consequently, the Auto-Bäcklund transformation (ABT) corresponding to the
governing PDE~(9) can be expressed as
\begin{equation}
	u=\frac{12\gamma}{\alpha}\frac{\partial^{2}}{\partial x^{2}}\bigl(\ln \psi\bigr)+u_{2},
\end{equation}
where $\psi(t,x,y)$ and $u_2(t,x,y)$ satisfy Eq.~(25) and Eq.~(9) respectively.
\newline
Furthermore, by choosing $u_2(t,x,y)=0$ in the above ABT, one obtains the
Cole--Hopf transformation in the form
\begin{equation}
	u=\frac{12\gamma}{\alpha}\frac{\partial^{2}}{\partial x^{2}}\bigl(\ln \psi\bigr).
\end{equation}
Hence, a variety of explicit solutions of the governing equation can be
constructed by selecting appropriate forms of $\psi$ and $u_2$.
\subsection{Exact solutions via ABT}
To obtain solitary wave solutions of the (1+2)-Dimensional Kudryashov-Sinelshchikov (KS) equation, we consider the following two cases.
\newline
\textbf{Case 1:} 
\begin{equation}
	\psi(t,x,y)=1+\exp\left(p(t)x+r(t)\right),
	\qquad
	u_2(t,x,y)=0,
\end{equation}
where $r(t)$, and $p(t)$ are arbitrary functions of time $t$. By substituting the above values of $\psi$ and $u_2$ into the Eq.~(25), we obtain
\begin{equation}
	\begin{aligned}
		&\exp\left(4xp(t)+4r(t)\right)
		\Biggl(
		-\frac{72\gamma e\,p(t)^4}{\alpha}
		-\frac{72\gamma x p(t)^3 p'(t)}{\alpha}
		-\frac{72\gamma p(t)^3 r'(t)}{\alpha}
		-\frac{72\gamma^{2} p(t)^6}{\alpha}
		\Biggr)=0, \\[1ex]
		&\exp\left(3xp(t)+3r(t)\right)
		\Biggl(
		\frac{144\gamma e\,p(t)^4}{\alpha}
		+\frac{144\gamma x p(t)^3 p'(t)}{\alpha}
		+\frac{72\gamma p(t)^2 p'(t)}{\alpha}
		+\frac{144\gamma p(t)^3 r'(t)}{\alpha}
		+\frac{144\gamma^{2} p(t)^6}{\alpha}
		\Biggr)=0, \\[1ex]
		&\exp\left(2xp(t)+2r(t)\right)
		\Biggl(
		-\frac{84\gamma e\,p(t)^4}{\alpha}
		-\frac{84\gamma x p(t)^3 p'(t)}{\alpha}
		-\frac{108\gamma p(t)^2 p'(t)}{\alpha}
		-\frac{84\gamma p(t)^3 r'(t)}{\alpha}
		-\frac{84\gamma^{2} p(t)^6}{\alpha}
		\Biggr)=0, \\[1ex]
		&\exp\left(xp(t)+r(t)\right)
		\Biggl(
		\frac{12\gamma e\,p(t)^4}{\alpha}
		+\frac{12\gamma x p(t)^3 p'(t)}{\alpha}
		+\frac{36\gamma p(t)^2 p'(t)}{\alpha}
		+\frac{12\gamma p(t)^3 r'(t)}{\alpha}
		+\frac{12\gamma^{2} p(t)^6}{\alpha}
		\Biggr)=0.
	\end{aligned}
\end{equation}
 by solving the above system of equations, we have
\begin{equation}
	p(t)=C_{1}, \qquad 
	r(t)=C_{2}-eC_{1}t-\gamma C_{1}^{3}t,
\end{equation}
where $C_{1}$ and $C_{2}$ are arbitrary integration constants. 
Substituting Eq.~(30) into Eq.~(28), we can see that it satisfied. Using Eq.~(26), the corresponding
analytical solution of the governing equation is derived as
\begin{equation}
	u_{A}(t,x,y)=
	\frac{12\gamma C_{1}^{2}
		\exp\left(-\gamma C_{1}^{3} t- C_{1} e t+ C_{1} x+ C_{2}\right)}
	{\alpha\left(1+
		\exp\left(-\gamma C_{1}^{3} t- C_{1} e t+ C_{1} x+ C_{2}\right)
		\right)}
	-
	\frac{12\gamma C_{1}^{2}
		\exp\left(-2\gamma C_{1}^{3} t-2 C_{1} e t+2 C_{1} x+2 C_{2}\right)}
	{\alpha\left(1+
		\exp\left(-\gamma C_{1}^{3} t- C_{1} e t+ C_{1} x+ C_{2}\right)
		\right)^{2}}.
\end{equation}
\newline
\textbf{Case 2:}
\begin{equation}
	\psi(t,x,y)=1+\exp(f(t)x+q(t)y+g(t)), 
	\qquad 
	u_2(t,x,y)=0,
\end{equation}
where $f(t)$, $q(t)$, and $g(t)$ are arbitrary functions of $t$.
\newline
Substituting $\psi$ and $u_2$ into Eq.~(25), we obtain 
\begin{equation}
	\begin{aligned}
		&-\frac{72\gamma f(t)^2}{\alpha}
		\exp\left(4(xf(t)+yq(t)+g(t))\right)
		\Big(
		e\,f(t)^2+\gamma f(t)^4+x\,f(t)f'(t)
		+d\,q(t)^2+y\,f(t)q'(t)+f(t)g'(t)
		\Big)=0, \\[1ex]
		&\frac{72\gamma f(t)^2}{\alpha}
		\exp\left(3(xf(t)+yq(t)+g(t))\right)
		\Big(
		f'(t)+2x\,f(t)f'(t)+2y\,f(t)q'(t)
		+2f(t)g'(t)+2e\,f(t)^2
		+2d\,q(t)^2+2\gamma f(t)^4
		\Big)=0, \\[1ex]
		&-\frac{12\gamma f(t)^2}{\alpha}
		\exp\left(2(xf(t)+yq(t)+g(t))\right)
		\Big(
		9f'(t)+7x\,f(t)f'(t)+7y\,f(t)q'(t)
		+7f(t)g'(t)+7e\,f(t)^2+7d\,q(t)^2
		\Big)\\
		& +7\gamma f(t)^4=0,\\[1ex]
		&\frac{12\gamma f(t)^2}{\alpha}
		\exp\left(xf(t)+yq(t)+g(t)\right)
		\Big(
		3f'(t)+x\,f(t)f'(t)+e\,f(t)^2
		+d\,q(t)^2+\gamma f(t)^4
		+y\,f(t)q'(t)+f(t)g'(t)
		\Big)=0.
	\end{aligned}
\end{equation}
Solving the above system, we obtain
\begin{equation}
	f(t)=D_{1}, 
	\qquad 
	q(t)=D_{2}, 
	\qquad
	g(t)=-\frac{dD_{2}^{2}}{D_{1}}t-\gamma D_{1}^{3}t-D_{1}et+D_{3},
\end{equation}
where $D_{1}$, $D_{2}$, and $D_{3}$ are arbitrary constants.
Substituting this result into Eq.~(32), the system is satisfied.
Using Eq.~(26), the corresponding analytical solution of the governing equation is
\begin{equation}
	\begin{aligned}
		u_{B}(t,x,y)=&
		\frac{12\gamma D_{1}^{2}
			\exp\left(D_{1}x+D_{2}y-\frac{dD_{2}^{2}}{D_{1}}t
			-\gamma D_{1}^{3}t-D_{1}et+D_{3}\right)}
		{\alpha\left(
			1+\exp\left(D_{1}x+D_{2}y-\frac{dD_{2}^{2}}{D_{1}}t
			-\gamma D_{1}^{3}t-D_{1}et+D_{3}\right)
			\right)} \\[1ex]
		&-
		\frac{12\gamma D_{1}^{2}
			\exp\left(2D_{1}x+2D_{2}y-\frac{2dD_{2}^{2}}{D_{1}}t
			-2\gamma D_{1}^{3}t-2D_{1}et+2D_{3}\right)}
		{\alpha\left(
			1+\exp\left(D_{1}x+D_{2}y-\frac{dD_{2}^{2}}{D_{1}}t
			-\gamma D_{1}^{3}t-D_{1}et+D_{3}\right)
			\right)^{2}}.
	\end{aligned}
\end{equation}
It is observed that two distinct analytical solutions, 
$u_{A}(t,x,y)$ and $u_{B}(t,x,y)$, as given in Eq.~(31) and Eq.~(35) have been successfully derived 
for the governing equation. To verify the validity of these solutions, 
they are substituted back into the original partial differential 
equation given in Eq.~(9). After direct substitution and appropriate 
simplification, both expressions satisfy the equation identically. 
This confirms that the obtained solutions represent exact analytical 
solutions of the considered $(1+2)$-dimensional 
Kudryashov--Sinelshchikov equation.
\subsection{Graphs and Physical Significance}
We present the two-dimensional and three-dimensional graphical representations of the exact solutions obtained in the previous section using the ABT method. These plots are generated by assigning specific parametric values to the variables involved, which allows a clear visualization of the behavior of the wave profile $u(t,x,y)$.
\begin{figure}[htbp]
	\centering
	\begin{subfigure}[b]{0.3\textwidth}
		\includegraphics[width=\textwidth]{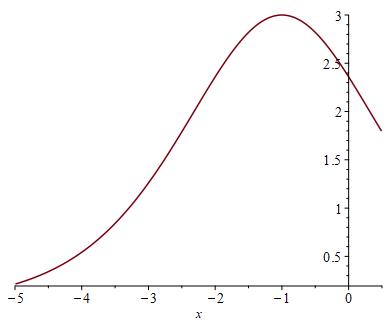}
		\caption{Two-dimensional plot}
	\end{subfigure}
	\hfill
	\begin{subfigure}[b]{0.4\textwidth}
		\includegraphics[width=\textwidth]{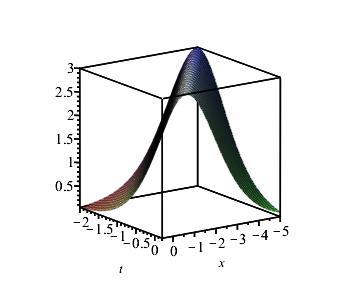}
		\caption{Three-dimensional surface plot}
	\end{subfigure}
	\caption{Graph of $u_{A}(t,x,y)$ corresponding to the values of 	$C_{1}=1$, $C_{2}=1$, $\alpha=1$, $\gamma=1$, $d=1$, and $e=1$.}
\end{figure}

\begin{figure}[htbp]
	\centering
	\begin{subfigure}[b]{0.4\textwidth}
		\includegraphics[width=\textwidth]{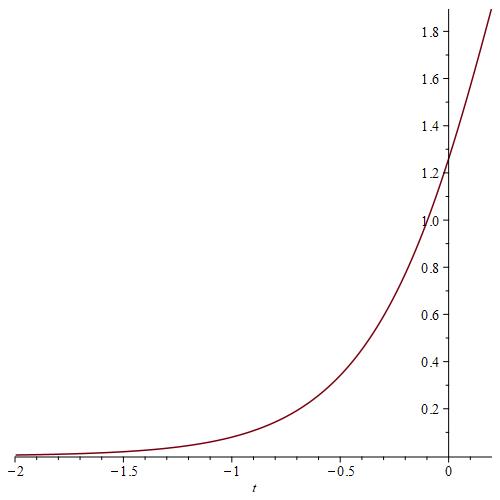}
		\caption{Two-dimensional plot}
	\end{subfigure}
	\hfill
	\begin{subfigure}[b]{0.4\textwidth}
		\includegraphics[width=\textwidth]{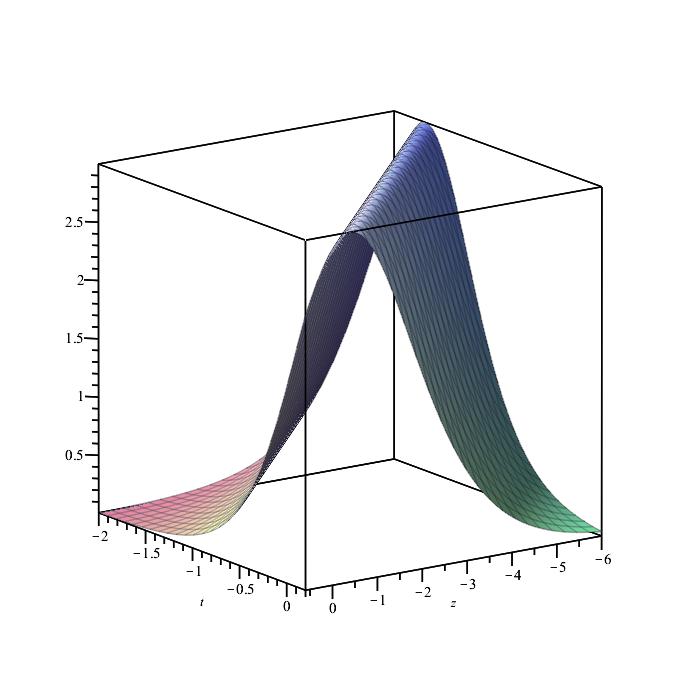}
		\caption{Three-dimensional surface plot}
	\end{subfigure}
	\caption{Graph of $u_{B}(t,x,y)$ corresponding to the values of 	$D_{1}=1,D_{2}=1$, $D_{3}=1$, $\alpha=1$, $\gamma=1$, $d=1$, and $e=1$.}
\end{figure}
Figure 1 illustrates the two-dimensional and three-dimensional graphical representations of the exact solution $u_A(t,x,y)$ corresponding to the parameter values $C_{1}=C_{2}=1$, $\alpha=\gamma=d=e=1$. The spatial variable $x$ varies in the interval $[-5,5]$, while $t \in [-2,2]$. The solution exhibits a smooth, bell-shaped bright solitary wave profile that is well localized in space and maintains its amplitude and shape during temporal evolution.
\smallskip

Figure 2 presents the two-dimensional and three-dimensional plots of the solution $u_B(t,x,y)$ for $D_{1}=D_{2}=D_{3}=1$, $\alpha=\gamma=d=e=1$. Here, the combined spatial variable $x+y$ varies in $[-6,6]$ with $t \in [-2,2]$. The obtained solution represents a stable bright solitary wave propagating obliquely in the $(x,y)$-plane, preserving its localized and non-singular structure throughout its evolution.
\section{Conclusion}
In this work, we investigated the $(1+2)$-dimensional Kudryashov–Sinelshchikov equation, which leads to the wave propagation of pressure waves through liquid consisting of gas bubbles. The essence of this study is the direct Painlev\'e analysis on PDE to investigate the integrability properties of the underlying nonlinear partial differential equation. We proved that the (1+2) KS equation is Painlev\'e integrable. 
\smallskip

In addition to this, the Auto-B\"acklund Transformation (ABT) was constructed by truncating the Laurent series expansion. 
The derived ABT was further used to solve the explicit analytical form of the nonlinear PDE by selecting suitable forms of the manifold function $\psi(t,x,y)$ in the form of exponents.  As a result, two representatives families of exact solutions were derived and directly verified to satisfy the original nonlinear PDE.
\smallskip

To better understand the physical characteristics of the obtained solutions, two-dimensional and three-dimensional graphical representations were presented. The plots reveal smooth, localized bright solitary wave structures that preserve their amplitude and shape during propagation. These results demonstrate the stable and physically meaningful behavior of the solutions and highlight the effectiveness of the direct Painlev\'e analysis of PDE combined with the ABT in deriving the exact solutions of the nonlinear evolution equations.
%

\section*{Acknowledgements}
RS and AP acknowledges the support of Anusandhan National Research Foundation (ANRF), India under the Core Research Grant with File No: CRG/2023/005418, dated 20 August, 2024.
\smallskip

This research is partially supported by the DST-FIST grant SR/FST/MS-I/2024/173 to the Department of Mathematics, Pondicherry University,
Puducherry, India.
\section*{Conflict of Interest}
The authors declare that there are no competing interests among the authors of this work.

\section*{Declaration}
Ethics, Consent to Participate, and Consent to Publish declarations: not applicable.

\end{document}